\documentclass[a4paper,UKenglish,ukenglish]{lipics-v2019} %

\usepackage{microtype}%

\usepackage{enumitem}
\usepackage{graphicx} %
\usepackage{tikz} 
\usepackage{wasysym}
\usetikzlibrary{arrows, automata, positioning,calc}
\usetikzlibrary{shapes,snakes}

\title{The Containment Problem for Unambiguous Register Automata}

\titlerunning{Containment for Unambiguous Register Automata}%

\author{Antoine Mottet}{Department of Algebra, Faculty of Mathematics and Physics, Charles University, Czech Republic}{}{https://orcid.org/0000-0002-3517-1745}{This author received funding from DFG Graduiertenkolleg 1763 (QuantLA) and from the European Research Council (ERC)
under the European Union’s Horizon 2020 research and innovation programme (grant agreement No 771005, “CoCoSym”).}%

\author{Karin Quaas}{University of Oldenburg, Germany}{}{}{Supported by DFG, QU~316/1-2.}

\authorrunning{A. Mottet and K. Quaas}%

\Copyright{Antoine Mottet and Karin Quaas}%



\category{}%

\relatedversion{}%

\supplement{}%

\funding{}%

\hideLIPIcs

\EventEditors{Rolf Niedermeier and Christophe Paul}
\EventNoEds{2}
\EventLongTitle{36th International Symposium on Theoretical Aspects of Computer Science (STACS 2019)}
\EventShortTitle{STACS 2019}
\EventAcronym{STACS}
\EventYear{2019}
\EventDate{March 13--16, 2019}
\EventLocation{Berlin, Germany}
\EventLogo{}
\SeriesVolume{126}
\ArticleNo{49}
\nolinenumbers %
\DeclareSymbolFont{extraup}{U}{zavm}{m}{n}
\DeclareMathSymbol{\varheart}{\mathalpha}{extraup}{86} %
\DeclareMathSymbol{\vardiamond}{\mathalpha}{extraup}{87} %

\newcommand\mN{\mathbb N}

\newcommand\ignore[1]{}

\DeclareMathOperator\dom{dom}

\newcommand{\tp}{\textup{tp}} %

\newcommand{\domain}{\mathbb{D}}
\newcommand{\data}{\textup{data}}
\newcommand{\proj}{\textup{proj}}

\newcommand{\A}{\mathcal{A}} 
\newcommand{\B}{\mathcal{B}} 
 
\newcommand{\locs}{\mathcal{L}}
\newcommand{\loc}{\ell}
\newcommand{\edges}{E}
\newcommand{\true}{\texttt{true}}
\newcommand{\init}{\textup{in}}
\newcommand{\acc}{\textup{acc}}

\newcommand{\config}{C}
\newcommand{\sconfig}{S} %
\DeclareMathOperator\abs{abs}
\newcommand{\sNodes}{\mathbb{S}} %
\newcommand{\sTo}{\Rightarrow} %

\renewcommand{\succ}{\textup{Succ}} 
\newcommand{\vect}[1]{\boldsymbol{#1}}

\newcommand{\indiscernible}[3]{{#1} \equiv_{#3} {#2}}

\newcommand{\register}{r}
\newcommand{\registers}{R}

\newcommand{\DRA}{\textup{DRA}}
\newcommand{\URA}{\textup{URA}}
\newcommand{\NRA}{\textup{NRA}}

\newcommand{\cDRA}{\mathbf{DRA}}
\newcommand{\cURA}{\mathbf{URA}}
\newcommand{\cNRA}{\mathbf{NRA}}

\newcommand{\regsim}{\sim}

\newcommand\EXPSPACE{\ensuremath{\textsf{EXPSPACE}}}
\newcommand\EEXPSPACE{\ensuremath{\textsf{2-EXPSPACE}}}
\newcommand\PSPACE{\ensuremath{\textsf{PSPACE}}} %

\begin{document}

\maketitle

\begin{abstract}
We investigate the complexity of the containment problem ``Does $L(\A)\subseteq L(\B)$ hold?'', where $\B$ is an unambiguous register automaton and $\A$ is an arbitrary register automaton.
We prove that the problem is decidable and give upper bounds on the computational complexity in the general case, and when $\B$ is restricted to have a fixed number of registers.
 \end{abstract}

\newpage
\section{Introduction}

Register automata~\cite{DBLP:journals/tcs/KaminskiF94} are a widely studied model of computation that extend finite automata with finitely many \emph{registers} that are able to hold values from
an infinite domain and perform equality comparisons with data from the input word.
This allows register automata to accept \emph{data languages}, i.e., sets of \emph{data words} over $\Sigma\times\domain$, where $\Sigma$ is a finite alphabet and $\domain$ is an infinite set called the data domain.
The study of register automata is motivated by problems in formal verification and database theory, where the objects under study are accompanied by annotations
(identification numbers, labels, parameters, ...), see the survey by S\'egoufin~\cite{DBLP:conf/csl/Segoufin06}.
One of the central problems in these areas is to check whether a given input document or program complies with a given input specification.
In our context, this problem can be formalized as a \emph{containment problem}: given two register automata $\A$ and $\B$, does $L(\A)\subseteq L(\B)$ hold, i.e.,  is the data language accepted by $\A$ included in the data language accepted by $\B$? Here, $\B$ is understood as a specification, 
and one wants to check whether $\A$ satisfies the specification.
For arbitrary register automata, the containment problem is undecidable~\cite{DBLP:journals/tocl/NevenSV04,DBLP:journals/tocl/DemriL09}.
It is known that one can recover decidability in two different ways.
First, the containment problem is known to be \textsf{PSPACE}-complete when $\B$ is a deterministic register automaton~\cite{DBLP:journals/tocl/DemriL09}.
This is a severe restriction on the expressive power of $\B$, and it is of practical interest to find natural classes of register automata that can be tackled algorithmically and that can express more properties than deterministic register automata.
Secondly, one can recover decidability of the containment problem when $\B$ is a non-deterministic register automaton with a single register~\cite{DBLP:journals/tcs/KaminskiF94,DBLP:journals/tocl/DemriL09}.
However, in this setting, the problem is Ackermann-complete~\cite{DBLP:conf/lics/FigueiraFSS11}; it can therefore hardly be considered tractable.

This motivates the study  of \emph{unambiguous} register automata, which are non-deterministic register automata for which every data word has at most one accepting run. Such automata are strictly more expressive than deterministic register automata~\cite{DBLP:journals/tcs/KaminskiF94,KaminskiZeitlin}. 

In the present paper, we investigate the complexity of the containment problem when $\B$ is restricted to be an unambiguous register automaton.
We prove that the problem is decidable with a \EEXPSPACE\  complexity,
and is even decidable in \EXPSPACE\ if the number of registers of $\B$ is a fixed constant. This is a striking difference to the non-deterministic case, where even for a fixed number of registers greater than $1$ the problem is undecidable. 
Classically, one way to approach the containment problem (for general models of computation) is to reduce it to a reachability problem on an infinite state transition system, called the \emph{synchronized state space of $\A$ and $\B$}, cf.~\cite{DBLP:conf/lics/OuaknineW04}.
Proving decidability or complexity upper bounds for the containment problem then amounts to finding criteria of termination or bounds on the complexity of a reachability algorithm on this space.
In this paper, our techniques also rely on the analysis of the synchronized state space of $\A$ and $\B$,
where our main contribution is to provide a bound on the size of synchronized states that one needs to explore before being able to certify that $L(\A)\subseteq L(\B)$ holds.
This bound is found by identifying elements of the synchronized state space whose behaviour is similar, and by showing that every element of the synchronized state space is equivalent to a small one.
In the general case, where $\B$ is unambiguous and $\A$ is an arbitrary non-deterministic register automaton, we bound the size of the graph that one needs to inspect by a triple exponential in the size of $\A$ and $\B$.
In the restricted case that $\B$ has a fixed number of registers, we proceed to give a better bound that is only doubly exponential in the size of $\A$ and $\B$.

\noindent
\subparagraph*{Related Literature}
A thorough study of the current literature on register automata reveals that there exists a variety of different definitions of register automata, partially with significantly different semantics. 
In this paper, we study register automata as originally introduced by Kaminski and Francez~\cite{DBLP:journals/tcs/KaminskiF94}. 
Such register automata process data words over an infinite data domain. 
The registers can take data values that appear in the input data word processed so far. The current input datum can be compared for (in)equality with the data that is stored in the registers. Kaminski and Francez study register automata mainly from a language-theoretic point of view; more results on the connection to logic, as well as the decidability status and computational complexity of classical decision problems like emptiness and containment are presented, e.g., 
in~\cite{DBLP:journals/tcs/SakamotoI00,DBLP:journals/tocl/NevenSV04,DBLP:journals/tocl/DemriL09}. 
In~\cite{DBLP:journals/corr/abs-1011-6432}, register automata over \emph{ordered} data domains are studied.

Kaminski and Zeitlin~\cite{KaminskiZeitlin} define a generalisation of the model in~\cite{DBLP:journals/tcs/KaminskiF94}, in the following called \emph{register automata with guessing}. 
The registers in such automata can non-deterministically reassign, or ``guess'', the datum of a register.
In particular, such register automata can store data values that have not appeared in the input data word before, in contrast to the register automata in~\cite{DBLP:journals/tcs/KaminskiF94}.
Register automata with guessing are strictly more expressive than register automata; for instance, there exists a register automaton with guessing that accepts the complement of the data language accepted by the register automaton in Figure \ref{fig:ura} (Example 4 in~\cite{KaminskiZeitlin}).  
Figueira~\cite{DBLP:journals/corr/abs-1202-3957} studies an alternating version of this model, also over ordered data domains. 
Colcombet~\cite{DBLP:conf/dcfs/Colcombet15,DBLP:conf/stacs/Colcombet12} considers \emph{unambiguous} register automata with guessing. 
In Theorem 12 in~\cite{DBLP:conf/dcfs/Colcombet15}, it is claimed that this automata class is effectively closed under complement, so that universality, containment and equivalence are decidable; however, to the best of our knowledge, this claim remains unproved.

Finally, unambiguity has become an important topic in automata theory, as witnessed by the growing body of literature in the recent years~\cite{DBLP:conf/concur/FijalkowR017, DBLP:conf/icalp/Skrzypczak18,DBLP:conf/icalp/DaviaudJLMP018,DBLP:conf/icalp/Raskin18}.
In addition to the motivations mentioned above, unambiguous automata form an important model of computation due to their \emph{succinctness} compared to their deterministic counterparts.
For example, it is known that unambiguous finite automata can be exponentially smaller than deterministic automata~\cite{DBLP:journals/ijfcs/Leung05}
while the fundamental problems (such as emptiness, universality, containment, equivalence) remain tractable.

\section{Main Definitions}

We study register automata as introduced in the seminal paper by Kaminski and Francez~\cite{DBLP:journals/tcs/KaminskiF94}. 
Throughout the paper, $\Sigma$ denotes a finite alphabet, and $\domain$ denotes an infinite set of data values.  In our examples, we assume $\domain=\mN$, the set of non-negative integers. 
A \emph{data word} is a finite sequence $(\sigma_1,d_1)\dots(\sigma_k,d_k) \in (\Sigma\times\domain)^*$. 
A \emph{data language} is a set of data words. 
We use $\varepsilon$ to denote the \emph{empty data word}. 
The \emph{length} $k$ of a data word $w$ is denoted by $|w|$. 
Given a data word $w$ as above and $0\leq i\leq k$, we define the infix $w(i,j]:=(\sigma_{i+1},d_{i+1})\dots(\sigma_j,d_j)$. Note that $w(i,i]=\varepsilon$.
We use $\data(w)$ to denote the set $\{d_1,\dots,d_k\}$ of all data occurring in $w$. We use 
$\proj(w)$ to denote the projection of $w$ onto $\Sigma^*$, i.e., the word $\sigma_1\dots\sigma_k$. 

Let $\domain_\bot$ denote the set $\domain\cup\{\bot\}$, where $\bot\not\in\domain$ is a fresh symbol not occurring in $\domain$.
A \emph{partial isomorphism} of $\domain_\bot$ is an injection $f\colon S\to\domain_\bot$ with finite domain $S\subset \domain_\bot$
such that if $\bot\in S$, then $f(\bot)=\bot$.
We use boldface lower-case letters like $\vect{a}, \vect{b}, \dots $ to denote tuples in $\domain_\bot^n$, where $n\in\mN$.
Given a tuple $\vect{a}\in\domain_\bot^n$, we write $a_i$ for its $i$-th component, and $\data(\vect{a})$ denotes the set $\{a_1,\dots,a_n\}\subseteq\domain_\bot$ of all data  occurring in $\vect{a}$.

Let $\registers=\{\register_1,\dots,\register_n\}$ be a finite set of \emph{registers}. 
A \emph{register valuation} is a mapping $\vect{a}:\registers\to\domain_\bot$; we may write $a_i$ as shorthand for $\vect{a}(\register_i)$. Let $\domain_\bot^\registers$ denote the set of all register valuations. 
Given $\lambda\subseteq\registers$ and $d\in\domain$, 
define the register valuation $\vect{a}[\lambda\leftarrow d]$ by 
$(\vect{a}[\lambda\leftarrow d])(\register_i):=d$ if $\register_i\in\lambda$, and $(\vect{a}[\lambda\leftarrow d])(\register_i):=a_i$ otherwise.

A \emph{register constraint} over $\registers$ is defined by the grammar 
\begin{align*}
\phi ::= \true \,\mid\, =\register \,\mid\, \neg\phi \,\mid\, 
  \phi \wedge \phi \, , 
\end{align*} 
where $\register\in\registers$. 
We use $\Phi(\registers)$ to denote the set of all register constraints over $\registers$. 
We may use $\neq\register$ or $\phi_1\vee\phi_2$ as shorthand for $\neg(=\register)$ and $\neg(\neg\phi_1\wedge\neg\phi_2)$, respectively. 
The satisfaction relation $\models$ for $\Phi(\registers)$ on $\domain_\bot^\registers\times\domain$ is defined by structural induction in the obvious way; e.g., $\vect{a},d\models (=\register_1 \, \wedge \, \neq\register_2)$ if $a_1=d$ and $a_2\neq d$.

A \emph{register automaton over $\Sigma$} is a tuple $\A=(\registers,\locs,\loc_\init,\locs_\acc,\edges)$, 
where 
\begin{itemize} 
\item $\registers$ is a finite set of registers, 
\item $\locs$ is a finite set of \emph{locations}, 
\item $\loc_\init\in\locs$ is the \emph{initial location}, 
\item $\locs_\acc\subseteq\locs$ is the set of \emph{accepting locations}, and
\item $\edges\subseteq \locs \times \Sigma \times \Phi(\registers) \times 2^\registers \times \locs$ is a finite set of \emph{edges}. We may write $\loc \xrightarrow{\sigma,\phi,\lambda}\loc'$ to denote an edge $(\loc,\sigma,\phi,\lambda,\loc')\in\edges$. Here, $\sigma$ is the label of the edge, $\phi$ is the register constraint of the edge, and $\lambda$ is the set of updated registers of the edge. A register constraint $\true$ is vacuously true and may be omitted; likewise we may omit $\lambda$ if $\lambda=\emptyset$. 
\end{itemize}
A \emph{state} of $\A$  is a pair $(\loc,\vect{a})\in \locs\times\domain_\bot^\registers$, where $\loc$ is the current location and $\vect{a}$ is the current register valuation. 
Given two states $(\loc,\vect{a})$ and $(\loc',\vect{a'})$ and some input letter $(\sigma,d)\in (\Sigma\times\domain)$, we postulate a transition $(\loc,\vect{a})\xrightarrow{\sigma,d}_\A(\loc',\vect{a'})$ if there exists some edge $\loc\xrightarrow{\sigma,\phi,\lambda}\loc'$ such that $\vect{a},d\models\phi$ and $\vect{a'}=\vect{a}[\lambda\leftarrow d]$. 
If the context is clear, we may omit the index $\A$ and write $(\loc,\vect{a})\xrightarrow{\sigma,d}(\loc',\vect{a'})$ instead of $(\loc,\vect{a})\xrightarrow{\sigma,d}_\A(\loc',\vect{a'})$. 
We use $\longrightarrow^*$ to denote the reflexive transitive closure of $\longrightarrow$.
A \emph{run} of $\A$ on the data word $(\sigma_1,d_1)\dots(\sigma_k,d_k)$ is a  sequence $(\loc_0,\vect{a^0}) \xrightarrow{\sigma_1,d_1} (\loc_1,\vect{a^1}) \xrightarrow{\sigma_2,d_2} \dots \xrightarrow{\sigma_k,d_k} (\loc_k,\vect{a^k})$ of transitions. 
We say that a run \emph{starts in $(\loc,\vect{a})$} if $(\loc_0,\vect{a^0})=(\loc,\vect{a})$. 
A run is \emph{initialized} if it starts in $(\loc_\init,\{\bot\}^\registers)$,  
and a run is \emph{accepting} if $\loc_k\in\locs_\acc$. 
The data language \emph{accepted} by $\A$, denoted by $L(\A)$, is the set of data words $w\in (\Sigma\times\domain)^*$ such that there exists an initialized  accepting run of $\A$ on $w$.

We classify register automata into \emph{deterministic register automata} (\DRA), \emph{unambiguous register automata} (\URA), and \emph{non-deterministic register automata} (\NRA).  
A register automaton is a \DRA \ if for every data word $w$ there is at most one initialized run. 
A register automaton is a \URA \ if for every data word $w$ there is at most one initialized accepting run. 
A register automaton without any restriction is an \NRA.
We say that a data language $L\subseteq (\Sigma\times\domain)^*$ is \DRA-recognizable (\URA-recognizable and \NRA-recognizable, respectively), if there exists a \DRA \ (\URA \ and \NRA, respectively) $\A$ over $\Sigma$ such that $L(\A)=L$. 
We write $\cDRA$, $\cURA$, and $\cNRA$ for the class of \DRA-recognizable, \URA-recognizable, and \NRA-recognizable, respectively, data languages. Note that $\cDRA \subseteq \cURA \subseteq \cNRA$.  Also note that, albeit a semantical property, the unambiguity of a register automaton can be decided using a simple extension of a product construction, cf.~\cite{DBLP:conf/dcfs/Colcombet15}.

The \emph{containment problem} is the following decision problem: given two register automata $\A$ and $\B$, does $L(\A)\subseteq L(\B)$ hold? 
We consider two more decision problems that stand in a close relation to the containment problem (namely, they both reduce to the containment problem): 
the \emph{universality problem} is the question whether $L(\B)=(\Sigma\times\domain)^*$ for a given register automaton $\B$. The \emph{equivalence problem} is to decide, given two register automata $\A$ and $\B$, whether $L(\A)=L(\B)$.

\section{Some Facts about Register Automata}
For many computational models, a straightforward approach to solve the containment problem is by a reduction to the emptiness problem using the equivalence: $L(\A)\subseteq L(\B)$ if, and only if, $L(\A)\cap \overline{L(\B)}=\emptyset$. 
This approach proves useful for $\cDRA$, which is closed under complementation. 
Using the decidability of the emptiness problem for $\NRA$, as well as the closure of $\cNRA$ under intersection~\cite{DBLP:journals/tcs/KaminskiF94}, 
we obtain the decidability of the containment problem for the case where $\A$ is an $\NRA$ and $\B$ is a $\DRA$. 
More precisely, and using results in~\cite{DBLP:journals/tocl/DemriL09}, the containment problem for this particular case is $\PSPACE$-complete. 

In contrast to $\cDRA$, the class $\cNRA$ is not closed under complementation~\cite{DBLP:journals/tcs/KaminskiF94} so that the above approach must fail if $\B$ is an $\NRA$. 
Indeed, it is well known that the containment problem for the case where $\B$ is an $\NRA$ is undecidable~\cite{DBLP:journals/tocl/DemriL09}. 
The proof is a reduction from the halting problem for Minsky machines: an $\NRA$ is capable to accept the complement of a set of data words encoding halting computations of a Minsky machine. 

\begin{figure}[t]
	\centering
  \scalebox{.8}{ 
\begin{tikzpicture}[->,>=stealth',shorten >=1pt,auto,node distance=4cm,thick,node/.style={circle,draw,scale=0.9}, roundnode/.style={circle, draw=black, thick, minimum size=6mm},]
\tikzset{every state/.style={minimum size=0pt}};
\node[roundnode,initial, initial text={}]  (1) at (0,0) {$\loc_0$};
\node[roundnode]  (2) at (2,0) {$\loc_1$};
\node[roundnode,accepting]  (3) at (4,0) {$\loc_2$};
\path [->] (1) edge node[above] {\scriptsize{$\{\register\}$}} (2);
\path [->] (2) edge node[above] {\scriptsize{$=\register$}} (3);
\path [->] (1) edge [loop above]  (1);
\path [->] (2) edge [loop above] node[above] {\scriptsize{$\neq\register$}} (2);
\node  (c) at (10,2) {$\{(\loc_0,\bot)\}$};
\node  (c1) at (8.5,1) {$\{(\loc_0,\bot),(\loc_1,1)\}$};
\node  (c2) at (11.5,1) {$\{(\loc_0,\bot),(\loc_1,2)\}$};
\node  (c11) at (8,0) {$\{(\loc_0,\bot),(\loc_1,1),(\loc_2,1)\}$};
\node  (c12) at (12,0) {$\{(\loc_0,\bot),(\loc_1,1),(\loc_1,2)\}$};
\node  (c111) at (7.5,-1) {$\{(\loc_0,\bot),(\loc_1,1),(\loc_2,1)\}$};
\node  (c112) at (9.8,-1) {$\dots$};
\node  (c123) at (12.5,-1) {$\{(\loc_0,\bot),(\loc_1,3),(\loc_1,2),(\loc_1,1)\}$};
\node  (c132) at (15.4,-1) {$\dots$};
\node  (c1234) at (12.5,-1.7) {$\dots$};
\node  (c1234) at (7.5,-1.7) {$\dots$};

\node  (c3) at (11.8,1.6) {$\dots$};
\path [->] (c) edge node[right] {\scriptsize{$\ 1$}} (c1);
\path [->] (c) edge node[left] {\scriptsize{$2 \ $}} (c2);
\path [->] (c1) edge node[right] {\scriptsize{$\ 1$}} (c11);
\path [->] (c11) edge node[right] {\scriptsize{$\ 1$}} (c111);
\path [->] (c2) edge node[left] {\scriptsize{$1 $}} (c12);
\path [->] (c12) edge node[right] {\scriptsize{$\ 3$}} (c123);

 	\end{tikzpicture}  
  }
\caption{
	On the left we depict a $\URA$ with a single register $\register$ and over a singleton alphabet (we omit the labels at the edges). The complement of the data language accepted by this $\URA$ cannot be accepted by any $\NRA$.  On the right we show a finite part of the infinite state space of the $\URA$. 
}
\label{fig:ura}
\end{figure}
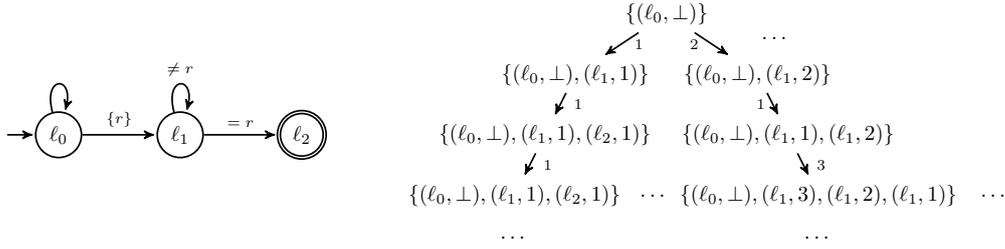

In this paper, we are interested in the containment problem for the case where $\A$ is an $\NRA$ and $\B$ is a $\URA$. 
When attempting to solve this problem, an obvious idea is to ask whether the class $\cURA$ is closed under complementation. Kaminski and Francez~\cite{DBLP:journals/tcs/KaminskiF94} proved that $\cURA$ is \emph{not} closed under complementation, and this even holds for the class of data languages that are accepted by $\URA$ that only use a single register. In Figure \ref{fig:ura}, we show a standard example of a $\URA$ for which the complement of the accepted data language cannot even be accepted by an $\NRA$~\cite{KaminskiZeitlin}.  Intuitively, this automaton is unambiguous because it is not possible for two different runs of the automaton on some data word to reach the location 
$\loc_1$  with the same register valuation at the same time. Therefore, at any time only one run can proceed to the accepting location $\loc_2$. 
Note that this also implies $\cDRA \subsetneq \cURA$.

An alternative approach for solving the containment problem is to explore the (possibly infinite) \emph{synchronized state space of $\A$ and $\B$}, cf.~\cite{DBLP:conf/lics/OuaknineW04}. 
Intuitively, the synchronized state space of $\A$ and $\B$ stores for every state $(\loc,\vect{a})$ that $\A$ is in after processing a data word $w$ the \emph{set of states} that $\B$ is in after processing the same data word $w$. 
For an example, see the computation tree on the right side of Figure \ref{fig:ura}, where the leftmost branch shows the set of states that the URA on the left side of Figure \ref{fig:ura} reaches after processing the data word $(\sigma,1)(\sigma,1)(\sigma,1)$, and the rightmost branch shows the set of states that the URA reaches after processing the data word $(\sigma,2)(\sigma,1)(\sigma,3)$. 
The key property of the synchronized state space of $\A$ and $\B$ is that it  contains sufficient information to decide whether for every data word for which there is an initialized accepting run in $\A$ there is also an initialized accepting run in $\B$. 
We formalize this intuition in the following paragraphs.

We start by defining the \emph{state space} of a given $\NRA$. 
Fix an \NRA \ $\A=(\registers,\locs,\loc_\init,\locs_\acc,\edges)$ over $\Sigma$. A \emph{configuration} of $\A$ is a finite set $\config\subseteq (\locs\times\domain^\registers_\bot)$ of states of $\A$; 
if $\config=\{(\loc,\vect{a})\}$ is a singleton set, in slight abuse of notation and if the context is clear, we may omit the parentheses and write $(\loc,\vect{a})$. 
Given a configuration $\config$ and an input letter $(\sigma,d)\in(\Sigma\times\domain)$, we use $\succ_\A(\config,(\sigma,d))$ to denote the \emph{successor configuration of $\config$ on the input $(\sigma, d)$}, formally defined by 
\begin{align*}
\succ_\A(\config,(\sigma,d)) := \{(\loc,\vect{a}) \in (\locs\times\domain^\registers_\bot) \mid \exists (\loc',\vect{a'})\in \config. (\loc',\vect{a'})\xrightarrow{\sigma,d}_\A(\loc,\vect{a})\}.\end{align*}
In order to extend this definition to data words, we define inductively $\succ_\A(\config,\varepsilon):=\config$ and $\succ_\A(\config, w \cdot (\sigma,d))  :=  \succ_\A(\succ_\A(\config,w),(\sigma,d))$. 
We say that a configuration $\config$ is \emph{reachable in $\A$} if there exists some data word $w$ such that $\config = \succ_\A((\loc_\init,\{\bot\}^\registers), w)$.  
We say that a configuration $\config$ is \emph{coverable in $\A$} if there exists some configuration $\config'\supseteq \config$ such that $\config'$ is reachable in $\A$. 
We say that a configuration $\config$ is \emph{accepting} if there exists $(\loc,\vect{a})\in\config$ such that $\loc\in\locs_\acc$; otherwise we say that $\config$ is \emph{non-accepting}. 
We define $\data(\config):=\bigcup_{(\loc,\vect{a})\in C}\data(\vect{a})$ as the set of data occurring in configuration $\config$.

The following proposition follows immediately from the definition of \URA.
\begin{proposition}
\label{prop:ura_implied_badness}
If $\A$ is a \URA \ and $\config,\config'$ are two configurations of $\A$  such that $\config\cap\config'=\emptyset$ and $\config\cup\config'$ is coverable, then for every data word $w$ the following holds: 
if $\succ_\A(\config,w)$ is accepting, then $\succ_\A(\config',w)$ is non-accepting. 
\end{proposition}

Let $\config,\config'$ be two configurations of $\A$. Consider two data words $w=(\sigma_1,d_1)\dots(\sigma_k,d_k)$ and $w'=(\sigma_1,d'_1)\dots(\sigma_k,d'_k)$   such that $\proj(w)=\proj(w')$.   
Recall that a partial function $f\colon \domain_\bot\to\domain_\bot$ with finite domain is a \emph{partial isomorphism} if it is an injection such that if $\bot\in\dom(f)$ then $f(\bot)=\bot$.
Let $f$ be a partial isomorphism of $\domain_\bot$ and let $\config$ be a configuration with $\data(\config)\subseteq \dom(f)$.
We define $f(\config):=\{(\loc,f(d_1),\dots, f(d_{|\registers|})) \mid  (\loc,d_1, \dots, d_{|\registers|})\in \config\}$; likewise, if $\{d_1,\dots,d_k\}\subseteq\dom(f)$,
we define $f((\sigma_1,d_1)\dots (\sigma_k,d_k)):=(\sigma_1,f(d_1))\dots (\sigma_k,f(d_k))$. 
We say that $\config,w$ and $\config',w'$ are \emph{equivalent with respect to $f$}, written $\config,w\regsim_f \config',w'$, if 
\begin{align}\tag{$\star$}
f(\config)=\config' \text{ and } f(w)=w'. \label{ra_equivalence_1} %
\end{align}
If $w=w'=\varepsilon$, then we may simply write $\config\regsim_f\config'$. 
We write $\config\regsim \config'$ if $\config\regsim_f \config'$ for some partial isomorphism $f$ of $\domain_\bot$.

\begin{proposition}
\label{prop:ra_equivalence}
If $\config,w \regsim \config',w'$, 
then $\succ_\A(\config,w(0,i]),w(i,k] \regsim \succ_\A(\config',w'(0,i]),w'(i,k]$ for all $0\leq i\leq k$, where $k=|w|$. 
\end{proposition}
\begin{proof}
The proof is by induction on $i$. 
For the induction base, let $i=0$. 
But then $\succ_\A(\config,w(0,0]))=\succ_\A(\config,\varepsilon)=\config$ and $w(0,k]=w$, and similarly for $\config'$ and $w'$, so that the statement holds by assumption. 
For the induction step, let $i>0$.  
Define $\config_{i-1}:=\succ_\A(\config,w(0,i-1])$ and similarly $\config'_{i-1}$. 
By induction hypothesis, there exists some bijective mapping $$f_{i-1}:\data(\config_{i-1})\cup \data(w(i-1,k]) \to \data(\config'_{i-1})\cup \data(w'(i-1,k])$$ satisfying  (\ref{ra_equivalence_1}) $f_{i-1}(\config_{i-1})=\config'_{i-1}$ and $f_{i-1}(w(i-1,k])=w'(i-1,k]$. 
Define $\config_i := \succ_\A(\config_{i-1},(\sigma_i,d_i))$ and $\config'_i:=\succ_\A(\config'_{i-1},(\sigma_i,d'_i))$. 
Note that $\data(\config_i)\subseteq\data(\config_{i-1})\cup\{d_i\}$, and similarly for $\data(\config'_i)$. 
Let $f_i$ be the restriction of $f_{i-1}$ to $\data(\config_i)\cup\data(w(i,k])$. 
We are going to prove that $\config_i,w(i,k] \regsim_{f_i} \config'_i,w'(i,k]$. 
Note that $f_i(w(i,k])=w'(i,k]$ holds by definition of $f_i$ and (2). 
We prove $f_i(\config_i)\subseteq \config'_i$. 
Suppose $(\loc,\vect{a})\in \config_i$. 
Hence there exists $(\loc_{i-1},\vect{b})\in \config_{i-1}$ such that $(\loc_{i-1},\vect{b})\xrightarrow{\sigma_i,d_i}(\loc,\vect{a})$. 
Thus there exists an edge $\loc_{i-1}\xrightarrow{\sigma_i,\phi,\lambda}\loc$  such that $\vect{b},d_i \models \phi$ and $\vect{a}=\vect{b}[\lambda\leftarrow d_i]$. 
By induction hypothesis, there exists $(\loc_{i-1},\vect{b'})\in\config'_{i-1}$ such that $f_{i-1}(\vect{b})=\vect{b'}$. 
By induction on the structure of $\phi$, one can easily prove that 
$\vect{b},d_i \models \phi$ if, and only if, $\vect{b'},d'_{i}\models\phi$. 
Define $\vect{a'}:=\vect{b'}[\lambda\leftarrow d'_i]$.  We prove $f_i(\vect{a})=\vect{a'}$: there are two cases: (i) If $\register\in\lambda$, then $f_i(\vect{a}(\register)) = f_i(d_i)=d'_i=\vect{a'}(r)$. (ii) If $\register\not\in\lambda$, then $f_i(\vect{a}(\register)) = f_i(\vect{b}(\register))=f_{i-1}(\vect{b}(\register))=\vect{a'}(r)$. 
Hence, $f_i(\vect{a})=\vect{a'}$. 
Altogether $(\loc, f_i(\vect{a}))\in\config'_i$, and thus $f_i(\config_i)\subseteq \config'_i$. The proof for $\config'_i\subseteq f_i(\config_i)$ is analogous. Altogether, $\config_i,w(i,k] \regsim_{f_i} \config'_i,w'(i,k]$. 
\end{proof}

As an immediate consequence of Proposition \ref{prop:ra_equivalence}, we obtain that $\sim$ preserves the configuration properties of being \emph{accepting} respectively \emph{non-accepting}. 
\begin{corollary}
\label{corollary:nra_bad}
Let $\config$ and $\config'$ be two configurations of $\A$.
If $\config,w\sim\config',w'$ and $\succ_\A(\config,w)$ is non-accepting (accepting, respectively), then $\succ_\A(\config',w')$ is non-accepting (accepting, respectively). 
\end{corollary}
Combining the last corollary with Proposition \ref{prop:ura_implied_badness}, we obtain 
\begin{corollary}
\label{corollary:ura_bad}
If $\A$ is a $\URA$ and $\config,\config'$ are two configurations such that $\config\cap\config'=\emptyset$ and $\config\cup\config'$ is coverable in $\A$, then 
for every data word $w$ such that $\config,w\sim\config',w$, the configurations 
$\succ_\A(\config,w)$ and $\succ_\A(\config',w)$ are non-accepting. 
\end{corollary}

For the rest of this paper, let  $\A=(\registers^\A,\locs^\A,\loc^\A_{\init},\locs^\A_{\acc},\edges^\A)$ be an \NRA \ over $\Sigma$, and let 
$\B=(\registers^\B,\locs^\B,\loc^\B_{\init},\locs^\B_{\acc},\edges^\B)$ be a \URA \ over $\Sigma$. 
Without loss of generality, we assume $\registers^\A \cap \registers^\B=\emptyset$ and $\locs^\A\cap\locs^\B=\emptyset$.  
We let $m$ be the number of registers of $\A$, and we let $n$ be the number of registers of $\B$.

A \emph{synchronized configuration of $\A$ and $\B$} is a pair 
$((\loc,\vect{d}),\config)$, where $(\loc,\vect{d})\in (\locs^\A\times\domain^{\registers^\A}_\bot)$ is a single state of $\A$, and
$\config\subseteq (\locs^\B\times\domain^{\registers^\B}_\bot)$ is a configuration of $\B$. 
Given a synchronized configuration $\sconfig$, we use $\data(\sconfig)$ to denote the set $\data(\vect{d}) \cup \data(\config)$ of all data occurring in $\sconfig$. 
We define $\sconfig_{\init}:=((\loc_{\init}^\A,\{\bot\}^m), \{(\loc_\init^\B,\{\bot\}^n)\})$ to be the \emph{initial synchronized configuration of $\A$ and $\B$}.
We define the \emph{synchronized state space of $\A$ and $\B$} to be the  (infinite) state transition system $(\sNodes,\sTo)$, where $\sNodes$ is the set of all synchronized configurations of $\A$ and $\B$, and $\sTo$ is defined as follows. If $\sconfig=((\loc,\vect{d}),\config)$ and $\sconfig'=((\loc',\vect{d'}),\config')$, then
$\sconfig\sTo\sconfig'$ if there exists a letter $(\sigma,d)\in(\Sigma\times\domain)$ such that $(\loc,\vect{d})\xrightarrow{\sigma,d}_\A(\loc',\vect{d'})$, and 
$\succ_\B(\config,(\sigma,d))=\config'$. 
We say that a synchronized configuration \emph{$\sconfig$ reaches a synchronized configuration $\sconfig'$ in $(\sNodes,\sTo)$} if there exists a path in $(\sNodes,\sTo)$ from $\sconfig$ to $\sconfig'$. 
We say that a synchronized configuration $\sconfig$ is \emph{reachable in $(\sNodes,\sTo)$} if $\sconfig_\init$ reaches $\sconfig$. 
We say that a synchronized configuration $\sconfig=((\loc,\vect{d}),\config)$ is \emph{coverable in $(\sNodes,\sTo)$} if there exists some synchronized configuration $\sconfig'=((\loc,\vect{d}),\config')$ such that $\config'\supseteq\config$ and $\sconfig'$ is reachable in $(\sNodes,\sTo)$.

We aim to reduce the containment problem $L(\A)\subseteq L(\B)$ to a reachability problem in $(\sNodes,\sTo)$. For this, call a  synchronized configuration $((\loc,\vect{d}),\config)$ \emph{bad} if $\loc\in\locs_\acc^\A$ is an accepting location and $\config$ is non-accepting, i.e., $\loc'\not\in\locs_\acc^\B$ for all $(\loc',\vect{a})\in\config$. 
The following proposition is easy to prove, cf.~\cite{DBLP:conf/lics/OuaknineW04}. 

\begin{proposition}
\label{prop:reductionToReach}
$L(\A)\subseteq L(\B)$ does not hold if, and only if, some bad synchronized configuration is reachable in $(\sNodes,\sTo)$. 
\end{proposition} 

We extend the equivalence relation $\sim$ defined above to synchronized configurations in a natural manner, i.e,  given a partial isomorphism $f$ of $\domain_\bot$
such that $\data(\vect{d})\cup\data(\config)\subseteq\dom(f)$,
we define $((\loc,\vect{d}),\config) \sim_f ((\loc,\vect{d}'),\config')$ if $f(\config)=\config'$ and $f(\vect{d})=\vect{d}'$.
We shortly write $\sconfig\sim\sconfig'$ if there exists a partial isomorphism $f$ of $\domain_\bot$ such that $\sconfig\sim_f\sconfig'$.
Clearly, an analogon of Proposition~\ref{prop:ra_equivalence} holds for this extended relation. In particular, we have the following:
\begin{proposition}\label{prop:equivalence-relation-synch-compatible}
Let $\sconfig,\sconfig'$ be two synchronized configurations of $(\sNodes,\sTo)$ such that $\sconfig\sim\sconfig'$.
If $\sconfig$ reaches a bad synchronized configuration, so does $\sconfig'$.
\end{proposition}

Note that the state transition system $(\sNodes,\sTo)$ is infinite. 
First of all,  $(\sNodes,\sTo)$ is not finitely branching: 
for every synchronized configuration  $\sconfig=((\loc,\vect{d}),\config)$ in $\sNodes$,  
every datum $d\in\domain$ may give rise to its own individual synchronized configuration $\sconfig_d$ such that $\sconfig\Rightarrow\sconfig_d$.  
However, it can be easily seen that for every two different data values $d,d'\in\domain\backslash\data(\sconfig)$, if inputting $(\sigma,d)$ gives rise to a transition $\sconfig\Rightarrow\sconfig_d$ and inputting $(\sigma, d')$ gives rise to a transition $\sconfig\Rightarrow\sconfig_{d'}$ (for some $\sigma\in\Sigma$), then $\sconfig_d\sim\sconfig_{d'}$. Hence there exist synchronized configurations $\sconfig_1, \dots, \sconfig_k$ for some $k\in\mN$ such that $\sconfig\sTo\sconfig_i$ for all $i\in\{1,\dots, k\}$, and such that for all $\sconfig'\in\sNodes$ with $\sconfig\sTo\sconfig'$ there exists $i\in\{1,\dots,k\}$ such that $\sconfig_i\sim\sconfig'$. 
This is why we define in Section~\ref{sect:abstract} the notion of \emph{abstract configuration}, representing synchronized configurations up to the relation $\sim$.
Second, and potentially more harmful for the termination of an algorithm to decide the reachability problem from Proposition \ref{prop:reductionToReach}, the configuration $\config$ of $\B$ in a synchronized configuration may grow unboundedly. As an example, consider the URA on the left side of Figure \ref{fig:ura}. For every $k\geq 1$, the configuration $\{(\loc_0,\bot),(\loc_1,d_1),(\loc_1,d_2)\dots,(\loc_1,d_k)\}$ with pairwise distinct data values $d_1,\dots,d_k$ is reachable in this URA by inputting the data word $(\sigma,d_1)(\sigma,d_2)\dots(\sigma,d_k)$.   
In the next section, we prove that we can solve the reachability problem from Proposition~\ref{prop:reductionToReach}
by focussing on a subset of configurations of $\B$ that are bounded in size, thus reducing to a reachability problem
on a finite graph.

\section{The Containment Problem for Register Automata}
\subsection{Types}

Given $k\in\mN$, a $k$-type\footnote{Types are a standard notion of model theory (see, e.g., \cite{Hodges} for a definition). The definition that we give here coincides with the standard notion of
types when applied to $\domain_\bot$.} of $\domain_\bot$ is a quantifier-free formula $\varphi(y_1,\dots,y_k)$ formed by a conjunction of (positive or negative) literals
of the form $y_i=y_j$ and $y_i=\bot$ that is satisfiable in $\domain_\bot$.
A $k$-type is \emph{complete} if for any other quantifier-free formula $\psi(y_1,\dots,y_k)$, either $\forall y_1,\dots, y_k.(\varphi(y_1,\dots,y_k)\Rightarrow\psi(y_1,\dots,y_k))$ holds or $\varphi\land\psi$ is
unsatisfiable.
It is easy to see that given $\vect{a}\in\domain^k$, there is a unique complete $k$-type $\varphi$ such that $\varphi(\vect a)$ holds in $\domain_\bot$.
We call $\varphi$ the \emph{type of $\vect{a}$} and denote it by $\tp(\vect{a})$.
It may be observed that $\vect a,\vect b\in\domain_\bot^k$ have the same type if, and only if, there exists a partial isomorphism $f$ of $\domain_\bot$
such that $f(\vect a)=\vect b$.

Recall that $m$ and $n$ denote the number of registers of $\A$ and $\B$. 
For every $\vect{a}\in\domain^n_\bot$ and for every complete $(2n+m)$-type $\varphi(\vect{y})$, where $\vect{y}=(y_1,\dots,y_{2n+m})$, 
we define the set \begin{align*}
\locs_\varphi(\vect{a})=\{ \loc'\in\locs^\B \mid \exists  \vect{b}\in \domain_\bot^n \text{ such that $(\loc',\vect b)\in C$ and }  \varphi(\vect{a},\vect{b},\vect{d}) \text{ holds in } \domain_\bot\}.\end{align*}  
Let $\sconfig=((\loc,\vect{d}),\config)$ be a synchronized configuration and let $\vect{a},\vect{b}\in\domain^n_\bot$ be two register valuations occurring in $\config$, i.e., there exist $\loc_{\vect a},\loc_{\vect b}\in\locs^\B$ such that $(\loc_{\vect a},\vect{a}),(\loc_{\vect b},\vect{b})\in\config$. 
We say that \emph{$\vect{a}$ and $\vect{b}$ are indistinguishable in $\sconfig$}, written $\indiscernible{\vect{a}}{\vect{b}}{\sconfig}$, 
if $\locs_\varphi(\vect{a}) = \locs_\varphi(\vect{b})$ for every complete $(2n+m)$-type $\varphi(\vect{y})$.

\begin{example}
\label{ex:types_and_indisc}
Let $(\loc^\A,3)$ be a state in some \NRA \ with a single register, and let $\config' = \{(\loc,1,3), (\loc,2,3), (\loc',1,2)\}$ be a configuration of a \URA \ with two registers. Let $\sconfig'=((\loc^\A,3),\config')$ be the corresponding synchronized configuration of $\A$ and $\B$. 
Consider $\vect{a}=(1,3)$ and $\vect{b}=(2,3)$.  
For the $5$-type
\begin{align*}\varphi_1 =  (y_1\neq y_2) \wedge (y_1 \neq y_3) \wedge (y_2=y_4) \wedge (y_4 =y_5)   \wedge (y_3\neq y_2)\end{align*} we have $\locs_{\varphi_1}(\vect{a})=\{\loc\}$ as $\varphi_1(\vect{a},\vect{b},\vect{d})$ holds in $(\mN,=)$, 
and similarly, 
$\locs_{\varphi_1}(\vect{b})=\{\loc\}$ as $\varphi_1(\vect{b},\vect{a},\vect{d})$ holds in $(\mN,=)$.  
However, we have $\locs_{\varphi_2}(\vect{a})=\{\loc'\}$ and $\locs_{\varphi_2}(\vect{b})=\emptyset$ for the $5$-type \begin{align*}\varphi_2 = (y_1\neq y_2) \wedge (y_1=y_3) \wedge (y_2\neq y_4) \wedge (y_2=y_5)  \wedge (y_4\neq y_1).\end{align*} Hence $\indiscernible{\vect{a}}{\vect{b}}{\sconfig'}$ does \emph{not} hold.  
However, $\indiscernible{\vect{a}}{\vect{b}}{\sconfig}$ for $\sconfig = ((\loc^\A,3),\config)$ with $\config:=\config' \cup \{(\loc',2,1)\}$.
\end{example}

\begin{proposition}
\label{prop:indiscernible_implies_regsim}
Let $\sconfig=((\loc^\A,\vect{d}), \config)$ be a coverable synchronized configuration of $\A$ and $\B$. Let  $\vect{a},\vect{b}$ be such that $\indiscernible{\vect{a}}{\vect{b}}{\sconfig}$. 
Then the map $f\colon\data(\vect{a})\to\data(\vect{b})$ defined by $f(a_i):=b_i$ is a partial isomorphism of $\domain_\bot$.
Moreover, if we let 
$\config_{\vect a} := \{(\loc,\vect{a})\in \config \mid \loc\in\locs^\B\}$ and 
$\config_{\vect b}:= \{(\loc,\vect{b})\in \config \mid \loc\in\locs^\B\}$,
then $\config_{\vect a} \regsim_f \config_{\vect b}$.
\end{proposition}
\begin{proof}
Let $\varphi$ be the complete $(2n+m)$-type of $(\vect{a},\vect{a},\vect{d})$.
Note that for two vectors $\vect{u},\vect{v}\in\domain^n_\bot$,  $\varphi(\vect{u},\vect{v},\vect{d})$ holds in $\domain_\bot$ iff $\vect{u}=\vect{v}$ and $\tp(\vect{a},\vect d)=\tp(\vect{u}, \vect d)=\tp(\vect{v}, \vect d)$.

Let now $(\loc,\vect a)$ be in $\config_{\vect a}$.
By definition, this means that $\loc\in\locs_\varphi(\vect{a})$. 
By indistinguishibility, $\loc\in\locs_\varphi(\vect{b})$ so that
\begin{align}\label{eq:indiscernible_implies_regsim}
\varphi(\vect{b},\vect{c},\vect{d}) \mbox{ holds in } \domain_\bot\tag{$\dagger$}
\end{align}
for some $(\loc,\vect{c})\in\config$.
Now, (\ref{eq:indiscernible_implies_regsim}) implies $\vect{b}=\vect{c}$ and $\tp(\vect b)=\tp(\vect a)$.
The former implies that $(\loc,\vect b)\in\config_{\vect b}$, while the latter implies that $f$ is a partial isomorphism.
Conversely, we obtain that $(\loc,\vect b)\in \config_{\vect b}$ implies $(\loc,\vect a)\in\config_{\vect a}$.
Hence $f(\config_{\vect a})=\config_{\vect b}$ and thus $\config_{\vect a}\regsim_f\config_{\vect b}$.
\end{proof}

\subsection{Collapsing Configurations}
As we pointed out in the introduction, 
the crucial ingredient of our algorithm for deciding whether $L(\A)\subseteq L(\B)$ holds is to prevent configurations $\config$ in a synchronized configuration $((\loc,\vect{d}),\config)$ to grow unboundedly. 
We do this by  \emph{collapsing two} subconfigurations $\config_{\vect a}, \config_{\vect b}\subseteq\config$ that behave equivalently with respect to reaching a bad synchronized configuration in $(\sNodes,\sTo)$ into a \emph{single} subconfiguration. 
The key notions for deciding when two subconfigurations can be collapsed into a single one are \emph{$k$-types} and \emph{indistinguishability} from the previous subsection.%

\begin{proposition}
\label{prop:collapse_URA}
Let $\sconfig'=((\loc,\vect{d}),\config')$ be a coverable synchronized  configuration of $\A$ and $\B$. 
	Let $\vect{a}$ and $\vect{b}$ be two distinct register valuations in $\config'$ such that $\indiscernible{\vect{a}}{\vect{b}}{\sconfig'}$.
	Let $\config_{\vect b} := \{ (\loc,\vect{b}) \in \config' \mid \loc\in\locs^\B \}$.
	Then $\sconfig:=((\loc,\vect{d}),\config'\setminus\config_{\vect b})$ reaches a bad synchronized configuration  if, and only if, $\sconfig'$ reaches a bad synchronized configuration. 
\end{proposition}
\begin{proof}
The ``if'' direction follows from the simple observation that for every data word $w$, if $\succ_\B(\config',w)$ is non-accepting, then so is $\succ_B(D,w)$ for every subset $D\subseteq \config'$.  
For the ``only if'' direction, 
let $\config_{\vect a} := \{ (\loc,\vect{a}) \in \config' \mid \loc\in\locs^\B \}$ and $\config :=\config'\setminus (\config_{\vect a}\cup \config_{\vect b})$.
Let $m$ be the number of registers of $\A$ and $n$ be the number of registers of $\B$.
Suppose that there exists a data word $w$  such that there exists an accepting run of $\A$ on $w$ that starts in $(\loc,\vect{d})$,  and
$\succ_\B(\config_{\vect a}\cup\config,w)$ is non-accepting. 
We assume in the following that $\succ_\B(\config_{\vect b},w)$ is accepting; otherwise  we are done. 
Without loss of generality, we assume that $\data(w) \cap \data(\sconfig')\subseteq \data(\vect b)\cup\data(\vect{d})$.
Otherwise, 
pick for every $d\in \data(w)\cap(\data(\vect a)\cup\data(\config))$ such that  $d\not\in\data(\vect b)\cup\data(\vect{d})$, a fresh datum $d'\in\domain$ not occurring in $\data(w)\cup\data(\sconfig')$, and simultaneously replace every occurrence of $d$ in $w$ by $d'$.  
	Let $w'$ be the resulting data word. 
	Then $(\loc,\vect{d}),w \regsim (\loc,\vect{d}),w'$ and $\config_{\vect b},w\regsim \config_{\vect b},w'$. 
	By Corollary \ref{corollary:nra_bad}, 
	$\succ_\A((\loc,\vect{d}),w')$ is accepting, and 
	$\succ_\B(\config_{\vect b},w')$ is accepting, too. 
	Then there must exist some accepting run of $\A$ on $w'$ starting in  $(\loc,\vect{d})$,  
	and, by Proposition \ref{prop:ura_implied_badness},  
	$\succ_\B(\config_{\vect a}\cup \config,w')$ must be non-accepting. 
	Hence, we could continue the proof with $w'$ instead of $w$.
	Let us assume henceforth that $\data(w) \cap \data(\sconfig')\subseteq\data(\vect b)\cup\data(\vect{d})$ holds.
	
	Let now $w''$ be the data word obtained from $w$ as follows: for every $b_i\in\data(w)$ with $b_i\neq a_i$, pick some fresh datum $e_i\in\domain$ not occurring in $\data(w)\cup\data(\sconfig')$. 
	Then replace every occurrence of the letter $b_i$ in $w$ by $e_i$. 
	
	Note that $(\loc,\vect{d}),w\sim(\loc,\vect{d}),w''$: 
	the key argument for this is that by $\indiscernible{\vect{a}}{\vect{b}}{\sconfig'}$ we have $b_i\not\in\data(\vect{d})$ whenever $b_i\neq a_i$. By Corollary \ref{corollary:nra_bad}, 
	$\succ_\A((\loc,\vect{d}),w'')$ is accepting. 
	Hence there must exist some accepting run of $\A$ on $w''$ starting in  $(\loc,\vect{d})$.  
	
	Further note that $\config_{\vect a},w''\regsim\config_{\vect b},w''$: 
	by Proposition \ref{prop:indiscernible_implies_regsim}, 
	$\config_{\vect a}\regsim_f\config_{\vect b}$, where
	$f:\data(\vect a)\to\data(\vect b)$ is the bijective mapping defined by 
	$f(a_i)=b_i$ for all $1\leq i \leq n$. 
	Now let $g:\data(\vect a)\cup\data(w'')\to\data(\vect b)\cup\data(w'')$ be the bijective mapping that agrees with $f$ on all data in $\data(\vect a)$, and that maps each datum $d\in\data(w'')\backslash\data(\vect a)$ to $d$. 
	One can easily see that $g$ is a bijection such that $g(\config_{\vect{a}})=\config_{\vect{b}}$ and $g(w'')=w''$ so that indeed $\config_{\vect a},w''\regsim_g\config_{\vect b},w''$. 
	By Corollary \ref{corollary:ura_bad}, $\succ_\B(\config_{\vect a},w'')$ and $\succ_\B(\config_{\vect b},w'')$ are non-accepting.

	Finally, we prove that $\succ_\B(C,w'')$ is non-accepting, too. 
	For this, let $(\loc',\vect{c})\in C$; we prove that $\succ_\B((\loc',\vect{c}),w'')$ is non-accepting. 
	We distinguish the following two cases:
	\begin{itemize}
	\item For all $1\leq i\leq n$ with $a_i\neq b_i$ we have  %
	$b_i\not\in\data(\vect{c})$. 
	Then $(\loc',\vect{c}),w \regsim (\loc',\vect{c}),w''$,  as witnessed by the bijection $f$ such that $f(b_i)=e_i$ for
	all $b_i\in\data(w)$ such that $b_i\neq a_i$, and that is the identity otherwise.
Recall that by assumption $\succ_\B((\loc',\vect{c}),w)$ is non-accepting. 
	By Corollary \ref{corollary:nra_bad}, 
	 $\succ_\B((\loc',\vect{c}),w'')$ is non-accepting. 
	\item There exists $1\leq i\leq n$ such that $a_i\neq b_i$ and $b_i\in\data(\vect{c})$. 

	Let $\varphi(\vect{y})$ be the $(2n+m)$-type of $(\vect{b},\vect{c},\vect{d})$, and note that $\loc'\in \locs_\varphi(\vect{b})$. 
	By assumption $\loc'\in\locs_\varphi(\vect{a})$ and there exists a state $(\loc',\vect{c'})\in \config$ such that
	$\varphi(\vect{a},\vect{c'},\vect{d})$ holds. 
	Note that for all $1\leq j\leq n$ such that $b_i=c_j$ we have $a_i=c'_j$. 
By assumption, $b_i = c_j$ for some $1\leq j\leq n$. 	
Since $a_i\neq b_i$, we can infer $c_j\neq c'_j$, and hence $(\loc',\vect{c})\neq(\loc',\vect{c'})$. %
	Next we prove 
	$(\loc',\vect{c}),w'' \regsim (\loc',\vect{c'}),w''$. 
	We define $f:\data(\vect{c})\cup\data(w'') \to \data(\vect{c'})\cup\data(w'')$
	as follows: 
	\begin{align*}
	f\colon \begin{cases} 
	c_p \mapsto c'_p & 1\leq p\leq n \\
	e \mapsto e & e\in\data(w'')
	\end{cases}
	\end{align*}
	We prove below that
	\begin{enumerate}[label=(\roman*)]
	\item for all $1\leq p,q\leq n$, $c_p=c_q$ iff $c'_p=c'_q$; 
	\item for all $1\leq p\leq n$, for all $e\in\data(w'')$,  $e=c_p$ iff $e=c'_p$;
	\end{enumerate}
	note that this implies that $f$ is well-defined and $f$ is a bijective mapping, and hence $(\loc',\vect{c}),w'' \regsim_f (\loc',\vect{c'}),w''$. 
	By Proposition \ref{prop:ra_equivalence}, 
	$\succ_\B((\loc',\vect{c}),w'') \regsim \succ_\B((\loc',\vect{c'}),w'')$. 
	By Corollary \ref{corollary:ura_bad},  $\succ_\B((\loc',\vect{c}),w'')$ and $\succ_\B((\loc',\vect{c'}),w'')$ are non-accepting.  
	We now prove the two items from above: 
	(i) Follows directly from the fact that $\varphi(\vect{a},\vect{c'},\vect{d})$ and $\varphi(\vect{b},\vect{c},\vect{d})$ hold,
	which implies that $\vect{c'}$ and $\vect{c}$ have the same type.
	For (ii), recall that $\data(w)\cap\data(\sconfig')\subseteq \data(\vect b)\cup\data(\vect{d})$. 
	This, the definition of $w''$, and $\indiscernible{\vect{a}}{\vect{b}}{\sconfig'}$ yield the claim. 
	\end{itemize}
	Altogether, 
	we proved that $\succ_\B(\config',w'')$ is non-accepting, 
	while there exists some accepting run $(\loc,\vect{d})\longrightarrow^*(\loc'',\vect{d''})$ of $\A$ on $w''$. This  finishes the proof for the ``only if'' direction. 
\end{proof}

When $\sconfig$ is obtained from $\sconfig'$ by applying Proposition~\ref{prop:collapse_URA} to some pair
of register valuations, we say that $\sconfig'$ \emph{collapses to} $\sconfig$.
We say that $\sconfig$ is \emph{maximally collapsed} if for all pairs $\vect{a}$ and $\vect{b}$ of distinct register valuations appearing in $\config$ we have that $\indiscernible{\vect{a}}{\vect{b}}{\sconfig}$ does \emph{not} hold.
Note that in Proposition~\ref{prop:collapse_URA}, the synchronized configuration $\sconfig$ is again coverable.
By iterating Proposition~\ref{prop:collapse_URA}, one obtains that a coverable synchronized configurations reaches a bad synchronized configuration
if, and only if, it collapses in finitely many steps to a maximally collapsed synchronized configuration that also reaches a bad synchronized configuration.

\begin{figure}[t]
	\centering
 \scalebox{.8}{ 
\begin{tikzpicture}[->,>=stealth',shorten >=1pt,auto,node distance=4cm,thick,node/.style={circle,draw,scale=0.9}, roundnode/.style={circle, draw=black, thick, minimum size=6mm},]
\tikzset{every state/.style={minimum size=0pt}};

\node at (-7.3,1) {\large{$\A$}};
\node[roundnode,initial, initial text={}]   (a1) at (-6.8,0) {};
\node[roundnode]   (a2) at (-5.6,0) {};
\node[roundnode]   (a3) at (-4.4,0) {};
\node[roundnode]   (a4) at (-3.2,0) {};
\node at (-3.2,0) {$\loc^\A$};
\node[roundnode,accepting]   (a5) at (-2,0) {};

\path [->] (a1) edge node[above] {\scriptsize{$\{\register\}$}} (a2);
\path [->] (a2) edge node[above] {\scriptsize{$\neq\register$}} (a3);
\path [->] (a3) edge node[above] {\scriptsize{$\{\register\}$}} (a4);
\path [->] (a4) edge (a5);

\node at (-0.5,1)  {\large{$\B$}};
\node[roundnode,initial, initial text={}]   (init) at (0,0) {};
\node[roundnode] (acc1) at (1.7,1) {};
\node[roundnode][below=1.7cm of acc1] (bad1) {};
\path [->] (init) edge[sloped,above] node[above] {\scriptsize{$\{\register_1 \}$}} (acc1);
\path [->] (init) edge[sloped,below] node[below] {\scriptsize{$\{\register_1 \}$}} (bad1);
\node[roundnode][right=1.5cm of acc1] (acc2) {};
\path [->] (acc1) edge node[above] {\scriptsize{$\neq\register_1, \{\register_2\}$}} (acc2);
\node[roundnode][right=1.5cm of bad1] (bad2) {};
\path [->] (bad1) edge[bend left=20] node[above] {\scriptsize{$\neq\register_1, \{\register_1\}$}} (bad2);
\path [->] (bad1) edge[bend right=20] node[below] {\scriptsize{$\neq\register_1$}} (bad2);
\node[roundnode][right=1.7cm of acc2] (acc3) {};
\node[right=-5.7mm of acc3] {$\loc'$};
\path [->] (acc2) edge  (acc3);
\node[roundnode][right=1.7cm of bad2] (bad3) {};
\node[right=-5.4mm of bad3] {$\loc$};
\path [->] (bad2) edge node[above] {\scriptsize{$ \{\register_2\}$}} (bad3);
\node[roundnode][below=0.5cm of acc3] (bad4) {};
\node[roundnode,accepting][right=2cm of bad4] (acc4) {};
\path [->] (acc4) edge [loop right]  (acc4);
\path [->] (acc3) edge node[left] {\scriptsize{$=\register_1\vee=\register_2$}} (bad4);
\path [->] (acc3) edge[sloped,above] node[above] {\scriptsize{$\neq\register_1 \wedge \neq\register_2$}} (acc4);
\path [->] (bad3) edge node[right] {\scriptsize{$\neq\register_1$}} (bad4);
\path [->] (bad3) edge[sloped,below] node[below] {\scriptsize{$=\register_1$}} (acc4);
	\end{tikzpicture}
  }
\caption{
An \NRA \ $\A$ and a \URA \ $\B$ over a singleton alphabet for which $L(\A)\subseteq L(\B)$.  
}
\label{fig:ura_no_collapse}
\end{figure}
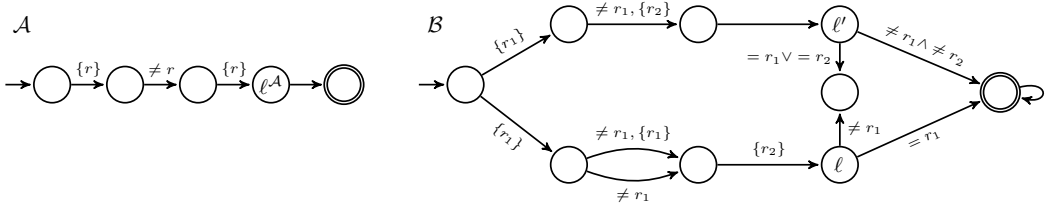
Before we present our algorithm for deciding the containment problem, we would like to point out that the intuitive notion of \emph{types} alone is not sufficient for deciding whether synchronized configurations can be collapsed. 
More precisely, given a coverable synchronized configuration $\sconfig'=((\loc^\A,\vect{d}),\config')$ and two register valuations $\vect{a}$ and $\vect{b}$ that occur in $\config'$ and for which $\tp(\vect{a},\vect{d})=\tp(\vect{b},\vect{d})$, 
it is in general \emph{not} the case that $\sconfig'$ reaches a bad synchronized configuration if $\sconfig:=((\loc,\vect{d}),\config'\backslash\config_{\vect{b}})$, where $\config_{\vect{b}}:=\{(\loc,\vect{b})\in\config'\mid\loc\in\locs^\B\}$, reaches a bad synchronized configuration. 
To see that, consider Figure \ref{fig:ura_no_collapse}, where two register automata over a singleton alphabet (we omit the labels at the edges) are depicted: an  \NRA \ $\A$ with a single register $\register$ on the left side, and
a \URA \ $\B$ with two registers $\register_1$ and $\register_2$ on the right side. Note that $L(\A)\subseteq L(\B)$. 
After processing the input data word $w=(\sigma,1)(\sigma,2)(\sigma,3)$, 
the synchronized configuration $\sconfig'=((\loc^\A,3),\config')$, where $\config':=\{(\loc,1,3),(\loc,2,3),(\loc',1,2)\})$, is reached in the synchronized state space of $\A$ and $\B$. For $\vect{a}=(1,3)$ and $\vect{b}=(2,3)$, we have $\tp(\vect{a},\vect{d})=\tp(\vect{b},\vect{d})$, but $\indiscernible{\vect{a}}{\vect{b}}{\sconfig'}$ does not hold (cf.\ Example \ref{ex:types_and_indisc}). Indeed, $\succ_\B(\config'\backslash\config_{\vect{b}},(\sigma,2))$ is non-accepting, while $\config'$ cannot reach any non-accepting configuration.  

\subsection{Abstract Configurations}\label{sect:abstract}
In this section, we study synchronized configurations up to the equivalence relation $\sim$.
Recall that $m$ is the number of registers of $\A$ and $n$ is the number of registers of $\B$.
	An \emph{abstract synchronized configuration of $\A$ and $\B$} is a tuple $(\loc, \Gamma, \varphi)$ where $\varphi$ is a complete $(sn+m)$-type for some $s\in\mN$,
	$\Gamma$ is an $s$-tuple of subsets of $\locs^\B$, and $\loc\in\locs^\A$.

The \emph{size} of an abstract synchronized configuration is defined to be $(sn+m)\log(sn+m) + s |\locs^\B| + \log(|\locs^\A|)$, which corresponds to the size needed on the tape of a Turing machine to encode an abstract synchronized configuration
(where one encodes, for example, an $(sn+m)$-type by giving for each of the $sn+m$ variables, a number in $\{1,\dots,sn+m\}$ in a way that $y_i=y_j$ is a conjunct in $\varphi$ iff $y_i$ and $y_j$ are assigned the same number).

Every synchronized configuration $\sconfig= ((\loc^\A,\vect d),\config)$ gives rise to an abstract synchronized configuration in the following way: let $\vect a^1,\dots,\vect a^s$ be the distinct register valuations in $\config$,
listed in some arbitrary order.
Let $\varphi$ be the complete $(sn+m)$-type of $(\vect a^1,\dots,\vect a^s,\vect d)$.
Let $C_{\vect a^i}:=\{ \loc \in \locs^\B \mid (\loc,\vect a^i)\in C\}$.
We obtain an abstract synchronized configuration $(\loc^\A, (C_{\vect a^1},\dots,C_{\vect a^s}), \varphi)$.
Different enumerations of the register valuations of $\config$ can yield different abstract configurations.
We let $\abs(\sconfig)$ be the set
of all abstract synchronized configurations that can be obtained from $\sconfig$. Every two abstract synchronized configurations in $\abs(\sconfig)$ can be obtained from one another by permuting the variables from the type
and the entries from the tuple accordingly.
It is easy to prove that $\sconfig\sim\sconfig'$ if, and only if, $\abs(\sconfig)=\abs(\sconfig')$.

An abstract configuration $(\loc,\Gamma,\varphi)$ is said to be \emph{maximally collapsed} if there exists a synchronized configuration $S$ such that $(\loc,\Gamma,\varphi)\in\abs(S)$ and such that $S$ is maximally collapsed (equivalently, one could ask that \emph{every} $S$ such that $(\loc,\Gamma,\varphi)\in\abs(S)$ is maximally collapsed).
The main result of this section is that the number of different register valuations in a maximally collapsed synchronized configuration is bounded.
Let $B_r\leq r^r$ be the number of complete $r$-types, which is also called the \emph{Bell number} of order $r$.

\begin{proposition}\label{prop:number_register_valuations}
	Let $\sconfig=((\loc^\A,\vect d),\config)$ be a maximally collapsed synchronized configuration of $\A$ and $\B$. The number of different register valuations appearing in $\config$
	is bounded by $(B_{2n+m}\cdot 2^{|\locs^\B|})^{(2n+m)^n}$.
\end{proposition}
\begin{proof}
We first prove a slightly worse upper bound, to give an idea of the proof.
Let $K:=B_{2n+m}$. We prove that the number of different register valuations is bounded by $2^{|\locs^\B|K}$.
Associate with every register valuation $\vect{a}$ appearing in $\config$ the $K$-tuple $(\mathcal{L}_{\varphi_1}(\vect{a}),\dots,\mathcal{L}_{\varphi_{K}}(\vect{a}))$ of subsets of $\locs^\B$,
where $\varphi_1,\dots,\varphi_{K}$ is an enumeration of all the complete $(2n+m)$-types.
Note that there are at most $2^{|\locs^\B|K}$ such tuples.
Suppose by contradiction that $\sconfig$ contains more than $2^{|\locs^\B|K}$ different register valuations. 
By the pigeonhole principle there are two different register valuations $\vect{a}$ and $\vect{b}$ that have the same associated $K$-tuple. 
Note that if $\vect{a}$ and $\vect{b}$ share the same $K$-tuple, then $\indiscernible{\vect{a}}{\vect{b}}{\sconfig}$. 
By Proposition~\ref{prop:collapse_URA}, $\sconfig$ could be collapsed further, contradiction.
Hence, we proved an upper bound of $2^{|\locs^\B|K}$ on the number of different register valuations appearing in a given maximally collapsed synchronized configuration.

We now proceed to prove the actual bound. The important fact is that when $\vect a$ and $\vect d$ are fixed in $\sconfig$,
then few (i.e., $\leq (2n+m)^n$) entries in the tuple $(\mathcal{L}_{\varphi_1}(\vect{a}),\dots,\mathcal{L}_{\varphi_{K}}(\vect{a}))$ are non-empty.
Indeed, in a given $(2n+m)$-type, each of the variables $y_{n+1},\dots,y_{2n}$ can be constrained to be equal to one of
$y_1,\dots,y_n,y_{2n+1},\dots,y_{2n+m}$, or constrained to be different than all of them.

Therefore, it remains to bound the number of $K$-tuples with entries in $2^{\locs^\B}$ and with at most $(2n+m)^n$ non-empty entries.
Each such tuple is characterised by the subset $T\subseteq\{1,\dots,K\}$ of entries that are non-empty, together with a $|T|$-tuple of non-empty subsets of $\locs^\B$. Since $|T|$ can be bounded by $(2n+m)^n$, we obtain that there are at most $K^{(2n+m)^n}\cdot 2^{|\locs^\B|(2n+m)^n}$
possible tuples, and thus at most $(B_{2n+m}\cdot 2^{|\locs^\B|})^{(2n+m)^n}$ different register valuations.
\end{proof}

Note that the bound in Proposition~\ref{prop:number_register_valuations} is doubly exponential in $n$ and exponential in $|\locs^\B|$ and $m$.
As a direct corollary, we obtain a bound on the number of maximally collapsed abstract synchronized configurations.
\begin{proposition}\label{prop:number_collapsed_config}
The number of maximally collapsed abstract configurations
is bounded by a triple exponential in $|\A|$ and $|\B|$.
If the number of registers of $\B$ is fixed, then the number of maximally collapsed abstract configurations is bounded by a double exponential in $|\A|$ and $|\B|$.
\end{proposition}
\begin{proof}
Recall that $m$ is the number of registers of $\A$ and $n$ is the number of registers of $\B$.
By Proposition~\ref{prop:number_register_valuations}, a maximally collapsed synchronized configuration $\sconfig=((\loc^\A,\vect d),\config)$
is such that $\config$ contains at most $K:=(B_{2n+m}\cdot 2^{|\locs^\B|})^{(2n+m)^n}$ different register valuations. Therefore, any abstract synchronized configuration in $\abs(\sconfig)$  is described by an $(sn+m)$-type
with $s\leq K$.
For a given $s$, there are at most $B_{sn+m}\cdot |\locs^\B|^s \cdot |\locs^\A|$ different abstract synchronized configurations.
Summing up from $s=0$ to $K$, we obtain that there are at most
\begin{align*}
\sum_{s=0}^{K} B_{sn+m}\cdot |\locs^\B|^s \cdot |\locs^\A| &\leq |\locs^\A|\cdot \left(B_m+B_{n+m}|\locs^\B|+\dots+ B_{nK+m}\cdot |\locs^\B|^{K}\right)\\
& \leq |\locs^\A| \cdot (1+K) \cdot B_{nK+m}\cdot |\locs^\B|^{K}\\
&\leq |\locs^\A| \cdot (1+K) \cdot (nK+m)^{(nK+m)}\cdot |\locs^\B|^{K}
\end{align*}
maximally collapsed abstract synchronized configurations. 
Since $K$ is doubly exponential in $|\A|$ and $|\B|$, this gives the first result. The second result follows from the fact that for fixed $n$, $K$
only depends exponentially on $m$ and $|\locs^\B|$.
\end{proof}

Given abstract synchronized configurations $(\loc^\A, \Gamma, \varphi)$ and $(\loc'^\A, \Gamma', \varphi')$,
define $(\loc^\A, \Gamma, \varphi)\leadsto (\loc'^\A, \Gamma', \varphi')$ if there exist synchronized configurations $\sconfig$ and $\sconfig'$
such that:
\begin{itemize}
\item $\sconfig\sTo\sconfig'$,
\item $(\loc^\A, \Gamma, \varphi)$ is in $\abs(\sconfig)$,
\item $\sconfig'$ can be maximally collapsed to some $\sconfig''$ such that $(\loc'^\A, \Gamma', \varphi')$ is in $\abs(S'')$.
\end{itemize}

\begin{lemma}\label{lem:abstract_relation_pspace}
	Given two abstract synchronized configurations $(\loc^\A, \Gamma, \varphi)$ and $(\loc'^\A, \Gamma', \varphi')$,
	deciding whether $(\loc^\A, \Gamma, \varphi)\leadsto (\loc'^\A, \Gamma', \varphi')$ holds can be done in polynomial space.
\end{lemma}
\begin{proof}
	In this proof, we assume without loss of generality that $\domain=\mN$.
	Let $s$ be such that $\varphi$ is an $(sn+m)$-type.
	Note that there is a synchronized configuration $\sconfig$ of the form $((\loc^\A,\vect d), D)$ such that $\data(D)\cup \data(\vect d)\subseteq\{1,\dots,sn+m\}$
	and such that $(\loc^\A, \Gamma, \varphi)\in\abs(\sconfig)$.
	This $\sconfig$ is moreover computable in polynomial space.
	
	To decide whether $(\loc^\A, \Gamma, \varphi)\leadsto (\loc'^\A, \Gamma', \varphi')$ holds, one simply:
	\begin{itemize}
		\item guesses a letter $\sigma\in\Sigma$ and a datum $d$ in $\{1,\dots,sn+m+1\}$,
		\item computes a synchronized configuration $\sconfig'$ obtained by firing the transition corresponding to $(\sigma,d)$ from $\sconfig$,
		\item guesses a sequence $(\vect a^1,\vect b^1),\dots,(\vect a^r,\vect b^r)$ of register valuations such that Proposition~\ref{prop:collapse_URA}
		can be applied $r$ times to obtain a maximally collapsed configuration $\sconfig''$,
		\item checks that $(\loc'^\A, \Gamma', \varphi')$ is in $\abs(\sconfig'')$. 
	\end{itemize}
	At the second step, the size of $\sconfig'$ is polynomially bounded by the size of $\A$, $\B$, and of $\sconfig$.
	Moreover, the maximal length of a collapsing sequence in the third step is also polynomially bounded, as the number of distinct register valuations
	decreases after each application of Proposition~\ref{prop:collapse_URA}.
	Therefore, this algorithm uses a polynomial amount of space.
\end{proof}

As for synchronized configuration, an abstract synchronized configuration $(\loc^\A, \Gamma, \varphi)$ is called \emph{bad} if $\loc^\A$ is an accepting location and none of the states in $\Gamma$ contains an accepting location.

\begin{proposition}\label{prop:correctness_abstraction}
	A bad synchronized configuration is reachable in $(\sNodes,\sTo)$ if, and only if, a bad abstract synchronized configuration is reachable from $\abs(\sconfig_\init)$.
\end{proposition}
\begin{proof}
	We prove that for every coverable synchronized configuration $S$ and every $n\geq 0$,
	a bad synchronized configuration is reachable in $n$ steps from $S$ if, and only if,
       a bad abstract synchronized configuration is reachable in $n$ steps from $\abs(S)$.
	The statement then follows by taking $S:=S_{\init}$.
	The proof goes by induction on $n$, where the case $n=0$ is trivial in both directions.
	
	Suppose now that $S$ reaches a bad synchronized configuration in $n$ steps.
	Let $S'$ be such that $S\sTo S'$ and such that $S'$ reaches a bad synchronized configuration in $n-1$ steps.
	Let $S''$ be such that $S'$ can be maximally collapsed to $S''$.
	By iterating Proposition~\ref{prop:collapse_URA}, we have that $S''$ reaches a bad synchronized configuration in $n-1$ steps
       (the fact that the length of the path is unchanged can be seen from the proof of Proposition~\ref{prop:collapse_URA}).
	It follows from the induction hypothesis that some $(\loc',\Gamma',\varphi')\in\abs(S'')$ reaches a bad abstract synchronized configuration in $n-1$ steps.
	Let $(\loc,\Gamma,\varphi)$ be an arbitrary abstraction in $\abs(S)$. We have by definition $(\loc,\Gamma,\varphi)\leadsto (\loc',\Gamma',\varphi')$,
	so that $(\loc,\Gamma,\varphi)$ reaches a bad abstract synchronized configuration in $n$ steps.
	The converse direction is proved similarly.
\end{proof}

Finally, we are able to present the main theorem.

\begin{theorem}\label{thm:main}
	The containment problem $L(\A)\subseteq L(\B)$, where $\A$ is a non-deterministic register automaton and $\B$ is an unambiguous register automaton, is in \EEXPSPACE.
	If the number of registers of $\B$ is fixed, the problem is in \EXPSPACE.
\end{theorem}
\begin{proof}
	The algorithm checks whether a bad abstract synchronized configuration is reachable from $\abs(\sconfig_{\init})$,
	       using the classical non-deterministic logspace algorithm for reachability.
       Every node of the graph can be stored using double-exponential space (see the second paragraph at the beginning of Section~\ref{sect:abstract}),
               and the size of the graph is triply exponential in the size of $\A$ and $\B$ by Proposition~\ref{prop:number_collapsed_config}.
  Moreover, the relation $\leadsto$ is decidable in polynomial space by Lemma~\ref{lem:abstract_relation_pspace}.
       Therefore, we obtain that the algorithm uses at most a double-exponential amount of space.
	In case the number of registers of $\B$ is fixed, Proposition~\ref{prop:number_collapsed_config}
	implies that the size of the graph is doubly exponential in the size of $\A$ and $\B$.
       We obtain that the algorithm uses at most an exponential amount of space.
\end{proof}

As an immediate corollary of Theorem~\ref{thm:main},
we obtain that the universality problem is in \EEXPSPACE\ and in \PSPACE\ for fixed number of registers.  Similarly, the equivalence problem for unambiguous register automata is in \EEXPSPACE.

\section{Open Problems}
The most obvious problem is to figure out the \emph{exact} computational complexity of the containment problem $L(\A)\subseteq L(\B)$, when $\B$ is an $\URA$. 
Finding lower bounds for unambiguous automata is a hard problem. 
Techniques for proving lower complexity bounds of the containment problem  (respectively the universality problem) for the case where $\B$ is a non-deterministic automaton rely heavily on non-determinism (cf. Theorem 5.2 in \cite{DBLP:journals/tocl/DemriL09}); 
as was already pointed out in~\cite{DBLP:conf/dcfs/Colcombet15}, we are lacking techniques for finding lower computational complexity bounds for the case where $\B$ is unambiguous, even for the class of finite automata. 
Concerning the upper bound, computer experiments revealed that maximally collapsed synchronized configurations seem to remain small.
Based on these experiments, we believe that the bound in Proposition~\ref{prop:number_register_valuations} is not optimal and can be improved to $O(2^{poly(n,m,|\locs^\B|)})$.
If this is correct, we would obtain an \EXPSPACE\ upper-bound for the general containment  problem.

We also would like to study to what extent our techniques can be used to solve the containment problem for other computation models.
In particular, we are interested in the following:
\begin{itemize}
	\item One can extend the definition of register automata to work over an ordered domain, where the register constraints
	are of the form $<r$ and $>r$. Proposition~\ref{prop:collapse_URA} turns out to be false in this setting,
	but it seems plausible that there exists a collapsibility notion that would work for this model.
	\item An automaton $\B$ is said to be $k$-ambiguous if it has at most $k$ accepting runs for every input data word,
	and polynomially ambiguous if the number of accepting runs for some input data word $w$ is bounded by $p(|w|)$ for some polynomial $p$. 
	Again, it is likely that simple modifications of Proposition~\ref{prop:collapse_URA} would give an algorithm
	for the containment problem for $k$-ambiguous register automata.
	\item Last but not least, we would like to point out that our techniques cannot directly be applied to the class of unambiguous register automata
	\emph{with guessing} which we mentioned in the introduction. Thus, the respective containment problem remains open for future research.
\end{itemize}


\begin{thebibliography}{10}

\bibitem{DBLP:conf/icalp/2018}
Ioannis Chatzigiannakis, Christos Kaklamanis, D{\'{a}}niel Marx, and Donald
  Sannella, editors.
\newblock {\em 45th International Colloquium on Automata, Languages, and
  Programming, {ICALP} 2018, July 9-13, 2018, Prague, Czech Republic}, volume
  107 of {\em LIPIcs}. Schloss Dagstuhl - Leibniz-Zentrum fuer Informatik,
  2018.
\newblock URL: \url{http://www.dagstuhl.de/dagpub/978-3-95977-076-7}.

\bibitem{DBLP:conf/stacs/Colcombet12}
Thomas Colcombet.
\newblock Forms of determinism for automata (invited talk).
\newblock In Christoph D{\"{u}}rr and Thomas Wilke, editors, {\em 29th
  International Symposium on Theoretical Aspects of Computer Science, {STACS}
  2012, February 29th - March 3rd, 2012, Paris, France}, volume~14 of {\em
  LIPIcs}, pages 1--23. Schloss Dagstuhl - Leibniz-Zentrum fuer Informatik,
  2012.
\newblock URL: \url{https://doi.org/10.4230/LIPIcs.STACS.2012.1}, \href
  {http://dx.doi.org/10.4230/LIPIcs.STACS.2012.1}
  {\path{doi:10.4230/LIPIcs.STACS.2012.1}}.

\bibitem{DBLP:conf/dcfs/Colcombet15}
Thomas Colcombet.
\newblock Unambiguity in automata theory.
\newblock In Jeffrey Shallit and Alexander Okhotin, editors, {\em Descriptional
  Complexity of Formal Systems - 17th International Workshop, {DCFS} 2015,
  Waterloo, ON, Canada, June 25-27, 2015. Proceedings}, volume 9118 of {\em
  Lecture Notes in Computer Science}, pages 3--18. Springer, 2015.
\newblock URL: \url{https://doi.org/10.1007/978-3-319-19225-3_1}, \href
  {http://dx.doi.org/10.1007/978-3-319-19225-3_1}
  {\path{doi:10.1007/978-3-319-19225-3_1}}.

\bibitem{DBLP:conf/icalp/DaviaudJLMP018}
Laure Daviaud, Marcin Jurdzinski, Ranko Lazic, Filip Mazowiecki, Guillermo~A.
  P{\'{e}}rez, and James Worrell.
\newblock When is containment decidable for probabilistic automata?
\newblock In Chatzigiannakis et~al. \cite{DBLP:conf/icalp/2018}, pages
  121:1--121:14.
\newblock URL: \url{https://doi.org/10.4230/LIPIcs.ICALP.2018.121}, \href
  {http://dx.doi.org/10.4230/LIPIcs.ICALP.2018.121}
  {\path{doi:10.4230/LIPIcs.ICALP.2018.121}}.

\bibitem{DBLP:journals/tocl/DemriL09}
St{\'{e}}phane Demri and Ranko Lazic.
\newblock {LTL} with the freeze quantifier and register automata.
\newblock {\em {ACM} Trans. Comput. Log.}, 10(3), 2009.
\newblock URL: \url{http://doi.acm.org/10.1145/1507244.1507246}, \href
  {http://dx.doi.org/10.1145/1507244.1507246}
  {\path{doi:10.1145/1507244.1507246}}.

\bibitem{DBLP:journals/corr/abs-1202-3957}
Diego Figueira.
\newblock Alternating register automata on finite words and trees.
\newblock {\em Logical Methods in Computer Science}, 8(1), 2012.
\newblock URL: \url{http://dx.doi.org/10.2168/LMCS-8(1:22)2012}, \href
  {http://dx.doi.org/10.2168/LMCS-8(1:22)2012}
  {\path{doi:10.2168/LMCS-8(1:22)2012}}.

\bibitem{DBLP:conf/lics/FigueiraFSS11}
Diego Figueira, Santiago Figueira, Sylvain Schmitz, and Philippe Schnoebelen.
\newblock Ackermannian and primitive-recursive bounds with dickson's lemma.
\newblock In {\em Proceedings of the 26th Annual {IEEE} Symposium on Logic in
  Computer Science, {LICS} 2011, June 21-24, 2011, Toronto, Ontario, Canada},
  pages 269--278. {IEEE} Computer Society, 2011.
\newblock URL: \url{http://dx.doi.org/10.1109/LICS.2011.39}, \href
  {http://dx.doi.org/10.1109/LICS.2011.39} {\path{doi:10.1109/LICS.2011.39}}.

\bibitem{DBLP:journals/corr/abs-1011-6432}
Diego Figueira, Piotr Hofman, and Slawomir Lasota.
\newblock Relating timed and register automata.
\newblock In Sibylle~B. Fr{\"{o}}schle and Frank~D. Valencia, editors, {\em
  Proceedings 17th International Workshop on Expressiveness in Concurrency,
  EXPRESS'10, Paris, France, August 30th, 2010.}, volume~41 of {\em {EPTCS}},
  pages 61--75, 2010.
\newblock URL: \url{http://dx.doi.org/10.4204/EPTCS.41.5}, \href
  {http://dx.doi.org/10.4204/EPTCS.41.5} {\path{doi:10.4204/EPTCS.41.5}}.

\bibitem{DBLP:conf/concur/FijalkowR017}
Nathana{\"{e}}l Fijalkow, Cristian Riveros, and James Worrell.
\newblock Probabilistic automata of bounded ambiguity.
\newblock In Roland Meyer and Uwe Nestmann, editors, {\em 28th International
  Conference on Concurrency Theory, {CONCUR} 2017, September 5-8, 2017, Berlin,
  Germany}, volume~85 of {\em LIPIcs}, pages 19:1--19:14. Schloss Dagstuhl -
  Leibniz-Zentrum fuer Informatik, 2017.
\newblock URL: \url{https://doi.org/10.4230/LIPIcs.CONCUR.2017.19}, \href
  {http://dx.doi.org/10.4230/LIPIcs.CONCUR.2017.19}
  {\path{doi:10.4230/LIPIcs.CONCUR.2017.19}}.

\bibitem{Hodges}
Wilfrid Hodges.
\newblock {\em A shorter model theory}.
\newblock Cambridge University Press, Cambridge, 1997.

\bibitem{DBLP:journals/tcs/KaminskiF94}
Michael Kaminski and Nissim Francez.
\newblock Finite-memory automata.
\newblock {\em Theor. Comput. Sci.}, 134(2):329--363, 1994.
\newblock URL: \url{https://doi.org/10.1016/0304-3975(94)90242-9}, \href
  {http://dx.doi.org/10.1016/0304-3975(94)90242-9}
  {\path{doi:10.1016/0304-3975(94)90242-9}}.

\bibitem{KaminskiZeitlin}
Michael Kaminski and Daniel Zeitlin.
\newblock Finite-memory automata with non-deterministic reassignment.
\newblock {\em International Journal of Foundations of Computer Science},
  {Volume 21, Issue 05}, 2010.

\bibitem{DBLP:journals/ijfcs/Leung05}
Hing Leung.
\newblock Descriptional complexity of nfa of different ambiguity.
\newblock {\em Int. J. Found. Comput. Sci.}, 16(5):975--984, 2005.
\newblock URL: \url{https://doi.org/10.1142/S0129054105003418}, \href
  {http://dx.doi.org/10.1142/S0129054105003418}
  {\path{doi:10.1142/S0129054105003418}}.

\bibitem{DBLP:conf/icalp/Skrzypczak18}
{Micha{\l} Skrzypczak}.
\newblock Unambiguous languages exhaust the index hierarchy.
\newblock In Chatzigiannakis et~al. \cite{DBLP:conf/icalp/2018}, pages
  140:1--140:14.
\newblock URL: \url{https://doi.org/10.4230/LIPIcs.ICALP.2018.140}, \href
  {http://dx.doi.org/10.4230/LIPIcs.ICALP.2018.140}
  {\path{doi:10.4230/LIPIcs.ICALP.2018.140}}.

\bibitem{DBLP:journals/tocl/NevenSV04}
Frank Neven, Thomas Schwentick, and Victor Vianu.
\newblock Finite state machines for strings over infinite alphabets.
\newblock {\em {ACM} Trans. Comput. Log.}, 5(3):403--435, 2004.
\newblock URL: \url{http://doi.acm.org/10.1145/1013560.1013562}, \href
  {http://dx.doi.org/10.1145/1013560.1013562}
  {\path{doi:10.1145/1013560.1013562}}.

\bibitem{DBLP:conf/lics/OuaknineW04}
Jo{\"{e}}l Ouaknine and James Worrell.
\newblock On the language inclusion problem for timed automata: Closing a
  decidability gap.
\newblock In {\em 19th {IEEE} Symposium on Logic in Computer Science {(LICS}
  2004), 14-17 July 2004, Turku, Finland, Proceedings}, pages 54--63. {IEEE}
  Computer Society, 2004.
\newblock URL: \url{https://doi.org/10.1109/LICS.2004.1319600}, \href
  {http://dx.doi.org/10.1109/LICS.2004.1319600}
  {\path{doi:10.1109/LICS.2004.1319600}}.

\bibitem{DBLP:conf/icalp/Raskin18}
Mikhail Raskin.
\newblock A superpolynomial lower bound for the size of non-deterministic
  complement of an unambiguous automaton.
\newblock In Chatzigiannakis et~al. \cite{DBLP:conf/icalp/2018}, pages
  138:1--138:11.
\newblock URL: \url{https://doi.org/10.4230/LIPIcs.ICALP.2018.138}, \href
  {http://dx.doi.org/10.4230/LIPIcs.ICALP.2018.138}
  {\path{doi:10.4230/LIPIcs.ICALP.2018.138}}.

\bibitem{DBLP:journals/tcs/SakamotoI00}
Hiroshi Sakamoto and Daisuke Ikeda.
\newblock Intractability of decision problems for finite-memory automata.
\newblock {\em Theor. Comput. Sci.}, 231(2):297--308, 2000.
\newblock URL: \url{https://doi.org/10.1016/S0304-3975(99)00105-X}, \href
  {http://dx.doi.org/10.1016/S0304-3975(99)00105-X}
  {\path{doi:10.1016/S0304-3975(99)00105-X}}.

\bibitem{DBLP:conf/csl/Segoufin06}
Luc Segoufin.
\newblock Automata and logics for words and trees over an infinite alphabet.
\newblock In Zolt{\'{a}}n {\'{E}}sik, editor, {\em Computer Science Logic, 20th
  International Workshop, {CSL} 2006, 15th Annual Conference of the EACSL,
  Szeged, Hungary, September 25-29, 2006, Proceedings}, volume 4207 of {\em
  Lecture Notes in Computer Science}, pages 41--57. Springer, 2006.
\newblock URL: \url{https://doi.org/10.1007/11874683\_3}, \href
  {http://dx.doi.org/10.1007/11874683\_3} {\path{doi:10.1007/11874683\_3}}.

\end{thebibliography}
\end{document}